\def\isarxiv{1}

\def\paperTitle{Sima 1.0: A Collaborative Multi-Agent Framework for Documentary Video Production}

\def\paperAuthor{
Zhao Song\thanks{\texttt{magic.linuxkde@gmail.com}. Work done while at Simons Institute, UC Berkeley.}  
}

\ifdefined\isarxiv
\documentclass[11pt]{article}
\usepackage[numbers]{natbib}
\else
\documentclass{article} 
\usepackage{iclr2026_conference,times}

\fi

\ifdefined\isarxiv

\usepackage{amsmath}
\usepackage{amsthm}
\usepackage{amssymb}
\usepackage{algorithm}
\usepackage{subfig}
\usepackage{algpseudocode}
\usepackage{graphicx}
\usepackage{grffile}
\usepackage{wrapfig,epsfig}
\usepackage{url}
\usepackage{xcolor}
\usepackage{epstopdf}

\usepackage{bbm}
\usepackage{dsfont}

\else 

\usepackage{hyperref}
\usepackage{url}

\fi

\ifdefined\isarxiv

\else

\usepackage{amsmath}
\usepackage{amsthm}
\usepackage{amssymb}
\usepackage{algorithm}
\usepackage{subfig}
\usepackage{algpseudocode}
\usepackage{graphicx}
\usepackage{grffile}
\usepackage{wrapfig,epsfig}
\usepackage{url}
\usepackage{xcolor}
\usepackage{epstopdf}

\usepackage{bbm}
\usepackage{dsfont}

\fi
 
\allowdisplaybreaks

\ifdefined\isarxiv

\usepackage{tikz}
\usepackage{hyperref}  
\hypersetup{colorlinks=true,citecolor=blue,linkcolor=blue} 
\usetikzlibrary{arrows}
\usepackage[margin=1in]{geometry}

\else
\fi
 
\graphicspath{{./figs/}}

\theoremstyle{plain}
\newtheorem{theorem}{Theorem}[section]

\newtheorem{definition}[theorem]{Definition}

\ifdefined\isarxiv

\else
\renewcommand\cite\citep
\fi

\begin{document}

\ifdefined\isarxiv

\date{}
\title{\paperTitle}
\author{\paperAuthor}

\else

\title{\paperTitle}

\author{Antiquus S.~Hippocampus, Natalia Cerebro \& Amelie P. Amygdale \thanks{ Use footnote for providing further information
about author (webpage, alternative address)---\emph{not} for acknowledging
funding agencies.  Funding acknowledgements go at the end of the paper.} \\
Department of Computer Science\\
Cranberry-Lemon University\\
Pittsburgh, PA 15213, USA \\
\texttt{\{hippo,brain,jen\}@cs.cranberry-lemon.edu} \\
\And
Ji Q. Ren \& Yevgeny LeNet \\
Department of Computational Neuroscience \\
University of the Witwatersrand \\
Joburg, South Africa \\
\texttt{\{robot,net\}@wits.ac.za} \\
\AND
Coauthor \\
Affiliation \\
Address \\
\texttt{email}
}

%

\newcommand{\fix}{\marginpar{FIX}}
\newcommand{\new}{\marginpar{NEW}}

\maketitle

\fi

\ifdefined\isarxiv
  \maketitle
  \begin{abstract}
    
Content creation for major video-sharing platforms demands significant manual labor, particularly for long-form documentary videos spanning one to two hours. In this work, we introduce Sima 1.0, a multi-agent system designed to optimize the weekly production pipeline for high-quality video generation. The framework partitions the production process into an 11-step pipeline distributed across a hybrid workforce. While foundational creative tasks and physical recording are executed by a human operator, time-intensive editing, caption refinement, and supplementary asset integration are delegated to specialized junior and senior-level AI agents. By systematizing tasks from script annotation to final asset exportation, Sima 1.0 significantly reduces the production workload, empowering a single creator to efficiently sustain a rigorous weekly publishing schedule.

  \end{abstract}


\else

\begin{abstract}

Content creation for major video-sharing platforms demands significant manual labor, particularly for long-form documentary videos spanning one to two hours. In this work, we introduce Sima 1.0, a multi-agent system designed to optimize the weekly production pipeline for high-quality video generation. The framework partitions the production process into an 11-step pipeline distributed across a hybrid workforce. While foundational creative tasks and physical recording are executed by a human operator, time-intensive editing, caption refinement, and supplementary asset integration are delegated to specialized junior and senior-level AI agents. By systematizing tasks from script annotation to final asset exportation, Sima 1.0 significantly reduces the production workload, empowering a single creator to efficiently sustain a rigorous weekly publishing schedule.

\end{abstract}

\fi



\section{Introduction}

In this work, we introduce a multi-agent system designed to streamline the weekly production of high-quality, long-form documentary videos. Typically spanning one to two hours in length, these videos cover a diverse array of topics, including immersive travelogues (e.g., Universal Studios Hollywood \cite{universal_studios_hollywood}, Legoland \cite{legoland_california}, Museums, and Zoos), in-depth movie video essays, corporate retrospectives of major technology companies (such as Google, Microsoft, and Nvidia), and comprehensive histories of premier annual events (like NeurIPS, the Mathematical Olympiad, and the World Cup). The primary objective of this framework is to optimize the content creation pipeline, reducing the production workload to a highly manageable level. Ultimately, this system empowers a single human operator, collaborating with multiple AI agents, to sustain a weekly publishing schedule for feature-length content.

We partition the entire video generation pipeline into a series of distinct stages. To optimize the workflow, tasks are distributed across a hybrid workforce: certain steps remain under the purview of a human operator, while others are delegated to either senior-level or junior-level AI agents depending on their complexity.

\begin{itemize}
    \item \textbf{Step 1. Determine the topic:} Select the subject matter for the video.
    \item \textbf{Step 2. Review source material:} Briefly watch relevant content (B-roll), categorized into two groups: public videos (Type A) and the author's original outdoor footage (Type B).
    \item \textbf{Step 3. Draft the script:} Write the complete script for the entire video.
    \item \textbf{Step 4. Record the A-roll:} Film the indoor segments, featuring the author presenting the script on camera.
    \item \textbf{Step 5. Polish the captions:} Generate and refine the subtitles for the indoor video recordings (A-roll).
    \item \textbf{Step 6. Source images:} Find relevant images---either public (Type A) or personal (Type B)---and annotate the script with precise insertion cues. 
    \item \textbf{Step 7. Source supplementary footage:} Find additional B-roll video clips and modify the script to indicate exactly when they should appear on screen.
    \item \textbf{Step 8. Collect B-roll materials}: Download B-roll videos and images 
    \item \textbf{Step 9. Edit the video:} Assemble the A-roll, images, and B-roll according to the script.
    \item \textbf{Step 10. Create transition graphics:} Design and insert title cards or visual transitions for the beginning of each of the video's segments (typically 15 parts).
    \item \textbf{Step 11. Finalize the project:} Review the complete video for pacing, accuracy, and quality assurance.
    \item \textbf{Step 12. Export the video:} Render and output the final video file.
\end{itemize}

We will dive into the details of every step.

{\bf Step 1. Determine the topic}

During the initial stages of a channel, it is beneficial to experiment with a wide variety of content. However, as the channel matures, it is optimal to narrow the focus to two or three core themes to establish a consistent audience.

\textbf{Step 2. Review source material}

Before detailing the proposed framework, we first establish key terminology foundational to video editing and production.

\begin{definition}[A-roll and B-roll, \cite{tb09}]
A-roll refers to the primary audio and visual footage that drives the main narrative of a video. It typically features the main subject directly addressing the camera or an interviewer, such as a "talking head" segment, a news anchor, or a scripted monologue. A-roll provides the foundational storyline and structural backbone of the edit.

B-roll refers to the supplementary visual footage intercut with the primary narrative footage (A-roll). Its primary functions are to provide visual context, illustrate the subject matter being discussed in the main-track audio, and conceal editorial transitions or cuts within the primary video timeline.
\end{definition}

We categorize the supplementary footage used in our system into two distinct types based on its origin.

\begin{definition}[B-roll, Type A, and Type B]
We define Type A as publicly available, free-to-use footage. We define Type B as original, outdoor footage filmed directly by the author of the channel. Both categories serve as supplementary video (B-roll) and will be referred to simply as Type A and Type B.
\end{definition}

The availability of Type B footage depends heavily on the documentary's subject matter. For location-based topics (e.g., visiting Universal Studios or Legoland), the author can easily record on-site Type B footage. However, for more abstract or historical topics, Type B footage may not be an option. In such cases, the video relies entirely on Type A footage combined with the primary main-track video.

\begin{definition}[A-roll, main-track video and splits]
The main-track video, commonly known in video production as A-roll, consists of indoor footage where the author presents the pre-written script. Due to the length of the content or production constraints, the main-track is often recorded in multiple sessions over several days. We refer to these individual, sequential recording sessions as splits (e.g., split A, split B, split C).
\end{definition}

\textbf{Step 3. Draft the script}

Typically, a script for a feature-length video is divided into 15 to 20 distinct sections. For illustrative purposes, assume a 15-part structure: Part 1 serves as the introduction, and Part 15 acts as the conclusion. The content and structure of the intervening sections (Parts 2 through 14) are highly adaptable and vary significantly depending on the specific topic being covered. 

To formalize the script structure, we define an additional key element:

\begin{definition}[Standard Call to Action (CTA), \cite{ee06}]
A Call to Action (CTA) is a standardized scripted segment designed to prompt specific audience engagement, such as explicitly instructing viewers to like the video or subscribe to the channel. 
\end{definition}

Within our framework, a CTA is embedded in two distinct locations in every video. While the core message is similar, the exact phrasing is context-dependent. The CTA integrated into Part 1 is designated as the Intro CTA, whereas the final prompt placed at the end of Part 15 is termed the  Concl CTA.

\textbf{Step 4. Record the A-roll}

This phase captures the primary narrative track, which is subsequently edited in DaVinci Resolve \cite{davinci_resolve}. Recording typically requires 80\% more time than the video's final runtime (a 1.8x multiplier). To manage operator fatigue, a standard 1.5-hour video is divided into three distinct recording sessions, referred to as splits. Split A covers Parts 1 through 5, Split B covers Parts 6 through 10, and Split C covers Parts 11 through 15. Consequently, generating 1.5 hours of final A-roll requires approximately 2.7 hours of active recording. Because this relies on the creator's physical presence, this step must be executed entirely by the human author.

\textbf{Step 5. Polish the captions}

Initially, an automated transcription tool generates baseline captions for the three main-track splits. However, these raw transcripts require a comprehensive polishing phase due to three persistent issues: 1) inherent speech-to-text anomalies, regardless of the speaker's clarity; 2) extended pauses or periods of silence during recording; and 3) redundant takes where the author repeatedly attempts a sentence due to prior errors. 

While this meticulous refinement has traditionally required human intervention, our framework delegates it entirely to a \textit{junior agent}. The efficiency gain is significant: a raw split intended to yield 0.5 hours of final content typically contains 0.9 hours of raw footage ($0.5\text{h} \times 1.8$). A human editor generally requires 150\% of the raw footage length to process and cut the video. Therefore, manually polishing the entire video would consume approximately 4.05 hours ($1.5\text{h} \times 1.8 \times 1.5$). By assigning this task to a junior agent, we eliminate a substantial manual workload.

\textbf{Step 6. Source images and annotate the script}

In this step, a human operator reviews the script sequentially to identify optimal insertion points for visual assets. The script is then annotated using a specific bracketing syntax, which varies based on the asset type:

\begin{itemize}
    \item \textbf{Type A (Public Assets):} The target text is wrapped in numerical tags. For example:
    \begin{center}
        This is the first sentence. \texttt{[1001+]} This is the second sentence. \texttt{[1001-]} This is the third sentence.
    \end{center}
    This notation indicates that image ID \texttt{1001} will be displayed concurrently with the narration of the second sentence. The spatial composition of the image depends on its hierarchical context within the script:
    \begin{itemize}
        \item \textbf{Standalone Layout:} If the image tags are independent (i.e., not nested within a video tag such as \texttt{[v1001+, 40s, 1.0x]}), the image is rendered full-screen over the main-track.
        \item \textbf{Nested Layout:} If the image tags are nested within an active video segment, the editing agent must determine the optimal compositional layout, choosing between a full-screen override or a localized overlay (picture-in-picture).
    \end{itemize}
    To maintain visual pacing, a single image should typically cover no more than three spoken sentences unless the asset is exceptionally detailed or engaging. 
    
    The syntax also supports simultaneous multi-image displays. If two images must appear during the same caption, they are concatenated within the tags:
    \begin{center}
        This is the first sentence. \texttt{[1001, 1002+]} This is the second sentence. \texttt{[1001, 1002-]} This is the third sentence.
    \end{center}
    Because Type A assets are publicly sourced, the operator must provide an accompanying comment specifying the direct URL source link for each referenced image.

    \item \textbf{Type B (Original Assets):} The target text is wrapped in descriptive, file-specific tags (e.g., \texttt{[image\_part0001+]} This is an example sentence. \texttt{[image\_part0001-]}). The display logic remains identical to Type A. However, because these assets are sourced internally from the author's local files, no external links or supplementary comments are required.
\end{itemize}

\textbf{Step 7. Source supplementary footage}

\begin{itemize}
    \item \textbf{Type A (Public Assets):} The target text is wrapped in tags prefixed with 'v' to denote video. For example:
    \begin{center}
        This is the first sentence. \texttt{[v1001+, 40s, 1.0x]} This is the second sentence. This is the third sentence. \texttt{[1001+]} This is the fourth sentence. \texttt{[1001-]} \texttt{[1003+]} This is the fifth sentence. \texttt{[1003-]} This is the sixth sentence. \texttt{[v1001-]} This is the seventh sentence.
    \end{center}
    This notation indicates that video ID \texttt{v1001} has a baseline duration of 40 seconds and is initially set to 1.0x playback speed. It is scheduled to cover the narrative from the second sentence through the sixth. If the video duration does not perfectly align with the text, the editing agent may dynamically extend its coverage (e.g., to include the first or seventh sentence) or adjust the playback speed (e.g., to 1.2x). If the speed is modified, the agent must update the script tag to reflect the new multiplier. Furthermore, the agent determines the compositional layout for the nested images (\texttt{1001} and \texttt{1003}), choosing whether to display them full-screen or as a localized overlay (picture-in-picture). Because Type A assets are publicly sourced, the opening tag \texttt{[v1001+, 40s, 1.0x]} must be accompanied by a comment specifying the source URL and precise extraction timestamps (e.g., \textit{Link URL, 11:10 - 11:50}).

    \item \textbf{Type B (Original Assets):} A similar syntax is applied using file-specific nomenclature:
    \begin{center}
        This is the first sentence. \texttt{[v\_part1001+, 1min, 1.0x]} This is the second sentence. This is the third sentence. \texttt{[1001+]} This is the fourth sentence. \texttt{[1001-]} \texttt{[image\_part1003+]} This is the fifth sentence. \texttt{[image\_part1003-]} This is the sixth sentence. \texttt{[v\_part1001-]} This is the seventh sentence.
    \end{center}
    This example demonstrates a complex nesting structure: one public image (\texttt{1001}) and one original image (\texttt{image\_part1003}) are inserted while a Type B video clip (\texttt{v\_part1001}) is actively playing. In this scenario, a source link comment is only required for the public Type A asset (\texttt{1001}).
\end{itemize}

As an illustrative example, consider a segment (e.g., Part 2) with a polished main-track duration of 10 minutes. The manual process of sourcing and annotating the script with all required supplementary B-roll materials—encompassing both Type A and Type B images and videos—requires 500\% additional time from a human operator. Consequently, a 10-minute segment necessitates 60 minutes of manual effort ($10 \times 6 = 60$ minutes).

 \textbf{Step 8. Collect B-roll materials}

Once the script is fully annotated with source links for the B-roll materials, an AI agent is deployed to automate the data collection process, entirely eliminating the need for manual downloading by a human operator. This extraction process is executed in five sequential steps:

\textbf{Step 8a: Install the Claude Code Extension.} Open VSCode, search for \textbf{Claude Code} in the Extensions marketplace, and install the plugin.
\begin{itemize} 
    \item If you possess an active Claude subscription, log in to your account and proceed directly to \textbf{Step 8d}. 
    \item If you do not have a Claude subscription, you must configure an alternative model via CC Switch as detailed in \textbf{Step 8b}.
\end{itemize}

\textbf{Step 8b: Install CC Switch (For Alternative Models).} Download and install the appropriate operating system release of CC Switch (an open-source routing tool) from the official repository: \url{https://github.com/farion1231/cc-switch/releases}.

\textbf{Step 8c: Configure the API Provider.} Open the CC Switch application, click ``Add Provider,'' and input your preferred model information alongside the corresponding API key. Note that CC Switch automatically rewrites the Claude Code configuration files, requiring no manual JSON modification. 

\textbf{Step 8d: Initialize the Workspace.} In VSCode, navigate to \textbf{File $\rightarrow$ Open Folder} and select the designated asset working directory (specifically, the root folder containing the \texttt{AGENTS.md} file).

\textbf{Step 8e: Task Submission and Automated Processing.} Within the Claude Code chat interface, input the asset extraction tasks using the standardized formatting syntax. 

For Image Tasks, use the following format:
\begin{center} 
\texttt{1001: https://example.com/image1.jpg \\
1002: https://example.com/image2.png}
 
\end{center}

For Video Tasks, use the following format:
\begin{center}
\texttt{v1001: https://youtu.be/XXXXX 0:10-0:35\\ 
v1002: https://youtu.be/YYYYY 1:20-2:05
}
\end{center}

\textbf{Automated Processing:} Upon submission, the agent autonomously executes the following operations:
\begin{itemize}
    \item Concurrently downloads all requested assets.
    \item Transcodes the video files into a QuickTime-compatible format (H.264 + AAC).
    \item Verifies file integrity by validating file sizes and exact durations.
\end{itemize}

\textbf{Step 9. Edit the video}

The video editing process is parallelized on a per-split basis. For instance, if three senior agents are available, each agent is assigned a single split (e.g., Split A) to process concurrently.

Within a given split, the agent first overlays all Type A and Type B supplementary videos (B-roll) onto the main-track video at their designated timestamps. The objective is to maximize visual coverage across the segment. To align the duration of the B-roll with the main-track narration, agents may apply time-remapping: excessively long clips can be sped up (e.g., 2x or, in extreme cases, up to 4x), while shorter clips can be slowed down (e.g., 0.75x or 0.5x), provided the slow-motion playback remains visually acceptable.

Following the video adjustments, all specified Type A and Type B still images are inserted into the timeline. If visual gaps remain over the main-track video after all speed adjustments and initial image insertions are complete, the agent must query for and integrate additional supplementary images to fill the remaining voids.

To maintain dynamic pacing and visual hierarchy, we define four specific scenarios where the visual behavior of the A-roll must be explicitly controlled:

\begin{itemize}
    \item \textbf{Full-Screen A-roll:} The A-roll occupies the entire screen without any supplementary B-roll overlay. This layout is mandatory for specific narrative segments, such as the Intro CTA and Concl CTA.
    \item \textbf{Transition Graphics:} When the author dictates the text of a transition graphic (which occurs at the start of Parts 2 through 15), the A-roll may either be displayed full-screen or completely hidden. The editing agent determines this display state based on the duration of the subsequent B-roll asset. If the upcoming B-roll video is excessively long, rather than drastically increasing its playback speed, the agent can initiate the B-roll early to cover the transition sequence, thereby obscuring the A-roll. Conversely, if the B-roll duration is manageable, the A-roll remains full-screen during the transition. Note that Part 1 (the introduction) does not utilize a transition graphic.
    \item \textbf{Hidden A-roll with Persistent Audio:} The visual A-roll is entirely obscured by Type A or Type B supplementary assets, while the main-track audio continues to play. This layout is applied to the entirety of Part 1, with the exception of the Intro CTA.
    \item \textbf{Stylized Overlay (Picture-in-Picture):} During the intervening sections (Parts 2 through 14), the A-roll is occasionally displayed as a stylized overlay, shaped either as a circle or a vertical rectangle ($height > width$). Within each recording split, the editing agent must identify four optimal locations for this effect, ensuring a minimum interval of three minutes between each occurrence. Both the circular and rectangular formats must be utilized across these four selected locations.
\end{itemize}

Recall that the standard script structure consists of 15 parts. Assuming a single part has a duration of 10 minutes, the editing duration varies based on asset availability. If the A-roll and all supplementary B-roll assets are pre-sourced and ready for insertion, a junior agent requires 150\% additional time to edit the sequence. Therefore, a 10-minute segment takes 25 minutes to edit ($10 \times 2.5 = 25$ minutes). Conversely, if the A-roll is prepared but numerous B-roll assets are missing, the agent must dynamically source new Type A materials online during the editing process. This active sourcing requires 200\% additional time, resulting in an editing time of 30 minutes per 10-minute segment ($10 \times 3 = 30$ minutes).

\textbf{Step 10. Create transition graphics}

To clearly demarcate distinct sections of the video, the system generates dedicated transition graphics (title cards). We use Canva \cite{canva} for this step. When creating a new transition, graphic, we duplicate the existing Canva project for the transition graphics, then rename the copy. We name transition graphic as f0200.png, f0300.png, $\cdots$ and f1500.png for part 2 to part 15. Each graphic features a solid black background, with the visual content constrained to the central 80\% of the screen to maintain a clean margin. Within this active content area, the layout is vertically divided: the upper 40\% displays the textual title or chapter heading, while the lower 60\% features a representative thematic image. These percentages are highly flexible and vary significantly depending on the overarching topic and the specific section of the video. Generating each transition graphic requires approximately 5 minutes of manual effort from the human operator. Usually such transition graphic has 1920 $\times$ 1080 pixels. 

\textbf{Step 11. Finalize the project}

This finalization stage encompasses the generation of crucial video metadata and supplementary visual assets. Specifically, the system must finalize: 1) a visual progress bar overlaid on the video timeline; 2) the end-of-video acknowledgments or credits sequence; 3) optimized text metadata, including the video title, search keywords, and a concise summary; and 4) customized video thumbnails. The thumbnails must be exported in three distinct aspect ratios: 16:9 (compressed to under 2MB), 4:3, and a 1:1 square format ($3000 \times 3000$ pixels). Additionally, a circular variant of the channel's logo must be uniformly embedded into every thumbnail asset.

\textbf{Step 12. Export the final assets}

In the final stage, DaVinci Resolve \cite{davinci_resolve} is utilized to export three distinct deliverables: the finalized video file, an isolated audio track (intended for alternative distribution, such as podcasts), and a standalone caption file (e.g., SRT or VTT). Although the primary subtitles are already burned directly into the video timeline, generating an independent caption file is critical. Many content hosting platforms rely on these standalone files to power their automated translation systems, enabling the video to be dynamically localized into multiple languages.

{\bf Remark.}
The framework presented in this paper is grounded in practical video production experience gathered over a six-month period (October 2025 to March 2026).

\ifdefined\isarxiv
\bibliographystyle{alpha}
\bibliography{ref}
\else
\bibliographystyle{iclr2026_conference}
\bibliography{ref}
\fi

\end{document}